\documentclass[review]{elsarticle}

\usepackage{natbib}
\usepackage{lineno,hyperref}

\usepackage{wrapfig}

\usepackage{amssymb,amsmath}
\usepackage{bm,upgreek}
\usepackage{mathrsfs}
\usepackage{graphics}
\usepackage{amsfonts}
\usepackage{color}
\usepackage{multirow}
\usepackage{floatpag}

\usepackage[symbol]{footmisc}

\floatpagestyle{plain}

\usepackage{lipsum}

\usepackage{makecell}

\graphicspath{{Fig/}}

\newcommand\rd{{\rm{d}}}
\newcommand{\bfm}[1]{{\rm\bf #1}}

\definecolor{myviolet}{RGB}{255,0,255}
\definecolor{darkgreen}{rgb}{0,0.5,0}
\definecolor{shade}{RGB}{176,176,176}

\journal{Elsevier}

\begin{document}

\begin{frontmatter}

\title{
Energy and morphology of martensite--twinned martensite interface in CuAlNi shape memory alloy: a phase-field study$^\dagger$
}

\author[IPPT]{Seyedshoja Amini}
\ead{samini@ippt.pan.pl}

\author[IPPT]{Mohsen Rezaee-Hajidehi\corref{cor1}}
\ead{mrezaee@ippt.pan.pl}

\author[IPPT]{Stanis{\l}aw Stupkiewicz}
\ead{sstupkie@ippt.pan.pl}

\cortext[cor1]{Corresponding author.}

\address[IPPT]{Institute of Fundamental Technological Research (IPPT), Polish Academy of Sciences,\\
Pawi\'nskiego 5B, 02-106 Warsaw, Poland.}

\begin{abstract}
Needle-like twins are observed experimentally within the transition layer at the martensite--twinned martensite interface. We utilize a phase-field approach to investigate this microstructure. Our goal is to simulate the morphology of the transition layer and to perform a detailed analysis to characterize its interfacial and elastic micro-strain energy. To illustrate the micromechanical framework developed for that purpose, sample computations are carried out for a CuAlNi shape memory alloy undergoing the cubic-to-orthorhombic martensitic transformation. A particular focus of the study is on size-dependent morphology through examining the impact of twin spacing. Additionally, our results reveal that certain twin volume fractions lead to the emergence of twin branching, as a way to minimize the total free energy stored in the microstructure.\footnotetext[2]{Published in \emph{Comput.\ Mater.\ Sci.}, 230 (2023), 112472, doi: 10.1016/j.commatsci.2023.112472.} 
\end{abstract}


\end{frontmatter}

\section{Introduction}
Pseudoelasticity and shape memory effect are the two most prominent features of shape memory alloys (SMAs). These features are inherent to the martensitic phase transformation and to the accompanying microstructures which encompass a rich array of interfaces across various spatial scales. Notably, the martensite--martensite (twin) interfaces, which are intrinsically coherent and free of (micro) stresses, stand out as the most ubiquitous type that form the primary constituent of the intricate microstructures at higher scales \citep{Bhattacharya2003}. Experimental investigations have shown that the martensitic transformation often proceeds by the evolution of nested laminated microstructures consisting of (quasi) periodic, planar and macroscopically sharp interfaces \citep{patoor1996micromechanical,abeyaratne1996kinetics}. These characteristics, indeed, serve as the backbone of the crystallographic theory of martensite which postulates that the interfaces are macroscopically compatible and stress-free \citep{Ball1987fine,Bhattacharya2003}. 

Nevertheless, local incompatibilities do exist and are primarily concentrated within thin \emph{transition layers} along the macroscopic interfaces. The local incompatibilities must be accommodated by elastic strains (referred to as `elastic micro strains') accompanied by micro stresses, as a result of which the transition layers develop a microstructured morphology \citep{knowles1981nature,chu1993hysteresis, abeyaratne1996kinetics,schryvers2001lattice,liu2003atomic}. A well-known example of a macroscopic interface is the habit plane that mediates the austenite and the domain of twinned martensite (note that the austenite and a single variant of martensite rarely form a compatible interface). The morphology of the corresponding transition layer and its energetic characteristics have been extensively investigated in the literature using various approaches, including analytical estimates based on simplified kinematics \citep{zhang2009energy}, shape optimization method \citep{maciejewski2005elastic,stupkiewicz2007low}, and phase-field modeling \citep{lei2010austenite,tuuma2016size,tuuma2016phase}. It is generally acknowledged that the morphology of a transition layer is driven by the material's propensity to minimize the total free energy and that its complex pattern is governed by the interplay between the elastic micro-strain energy and the interfacial (surface) energy of the phase boundaries. Branching, i.e., refinement of twin spacing, in the vicinity of the macroscopic interface represents a well documented manifestation of morphological changes within the transition layers \citep{kohn1992branching, liu2003atomic,james2005way,seiner2008shape,seiner2020branching}.

\begin{figure}
\centering
\includegraphics[width=1\textwidth]{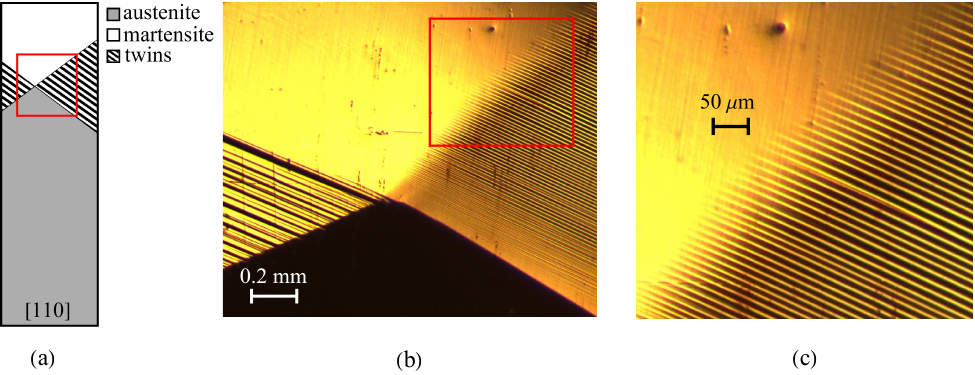}
\caption{$\lambda$-microstructure in CuAlNi: (a) a sketch of the macroscopic morphology, (b) a close-up view of the $\lambda$-microstructure at the intersection point, and (c) a close-up view of the martensite--twined martensite interface involving needles. The optical micrographs in panels (b,c) are provided courtesy of H. Seiner, see \citep{seiner2011finite} (reproduced with permission from Elsevier).}
\label{fig-seiner}
\end{figure}

At the same time, morphologies featuring needle-like twins emerge at the macroscopic interface between a single variant of martensite and twinned martensite or between two distinctly oriented twinned martensite domains \citep{chu1993hysteresis,schryvers2001lattice,seiner2008shape, seiner2011finite}. The so-called $\lambda$-microstructure is an example of a microstructure involving such macroscopic interfaces. As shown by Seiner et al.\ \cite{seiner2008shape,seiner2011finite}, a macroscopically non-uniform martensitic transformation is induced and controlled by a temperature gradient in a CuAlNi single crystal bar, and this leads to the formation of the $\lambda$-microstructure. This microstructure comprises four interfaces, all intersecting at one point, namely two austenite--twinned martensite and two martensite--twinned martensite interfaces (the latter referred to as "twinned-to-detwinned interface" by the authors). An optical micrograph of the resulting $\lambda$-microstructure is depicted in Fig.~\ref{fig-seiner}. A closer look at the corresponding martensite--twinned martensite interface, see Fig.1(c), reveals the needle-like appearance of the twins. The authors examined the structure of the needles via white-light interferometry and found out that the needles bend and taper as they approach the domain of pure martensite variant and that, at some locations, branching of the twins takes place. These distinctive characteristics have been also observed for a different SMA material, namely In-Tl \citep{basinski1954experiments}, and also at the macroscopic interface between two twinned martensite domains, e.g., \citep{chu1993hysteresis,schryvers2001lattice}. It is worth noting that needles, in general, have been reported in a variety of microstructures, e.g., \citep{salje1998needle,niemann2017nucleation, bronstein2019analysis,lauhoff2022effects}, and that a particular attention in the literature has been devoted to the theoretical and numerical analysis of needle-like morphologies, e.g., {\citep{li1999theory,li2001computations,finel2010phase,seiner2011finite, uchimali2021modeling,conti2020geometry,conti2023microstructure}}. {Within this context, a number of modeling approaches have demonstrated their potential in simulating complex spatially-resolved microstructures at the meso-scale, including the phase-field method \citep{finel2010phase} and the sharp-interface discrete-particle method \citep{uchimali2021modeling}.}

To understand the formation mechanism of intriguing macroscopic interfaces, it is necessary to analyze the morphology and to determine the energy-based characteristics of the transition layers. With these two objectives in mind, a detailed modeling-based investigation of the martensite--twinned martensite interface in a CuAlNi single crystal is pursued in this study, with a special emphasis on the related size effects. To accomplish this, we leverage a conventional (two-phase) phase-field model which has been simply derived from our earlier multiphase-field model \citep{tuuma2021phase}, and thereby retains its essential features, especially the finite-strain kinematics, consideration of the full elastic anisotropy of martensite variants and formulation in the variational framework. It should be stressed that the viscous evolution amounts to the minimization of the total free energy, comprising the elastic strain energy and the interfacial energy, thus making the phase-field method a suitable framework to address the problem at hand. {It is noteworthy that the primary focus of our analysis is on the transition layers with the needle of one variant \emph{not} terminating at the same variant, as shown in Fig.~\ref{fig-seiner}(c).} To the best of our knowledge, the only closely-related study in this context is that by Seiner et al.\ \citep{seiner2011finite}, who examined the structure of {such} needles via a finite-element-based sharp-interface model.

A well-known drawback of the phase-field method is its requirement for a sufficiently dense finite-element mesh in order to accurately resolve the diffuse interfaces, and thereby, to properly describe the associated interfacial energy \citep{levitas2011phase,tuuma2021phase}. At the same time, it is necessary to adopt a physically relevant value for the interfacial energy density {parameter}, which sets the length scale of our diffuse interfaces to the order of few nanometers. These two factors together limit the size of the computational domain that can be simulated (in our case, the computational domain is assumed periodic and encloses one twin pair). This size (on the scale of $<200$ nm) is visibly smaller than what has been observed in the experiment, which is of the order of 10 $\mu$m, see Fig.~\ref{fig-seiner}(c). {Therefore, while some qualitative comparisons have been drawn throughout the analysis, we do not aim for any direct quantitative comparison with the available experimental data.}

The paper is organized as follows. In Section~\ref{sec-crys}, we recall the basic equations of the crystallographic theory of martensite, in order to lay the theoretical foundation for the problem at hand. The phase-field model is briefly described in Section~\ref{sec-PF}. Subsequently, Section~\ref{sec-res} presents the simulation setup, the obtained results and the ensuing discussions.

\section{Basic equations of the crystallographic theory of martensite}\label{sec-crys}
According to the crystallographic theory of martensite, the requirement of kinematic compatibility is imposed at zero stress and implies that the deformation gradients on the opposite faces of a planar interface are rank-one connected \citep{Ball1987fine,Bhattacharya2003}. In line with this geometrical definition, the kinematic compatibility condition between two stress-free variants of martensite, here variant A and variant B, is mathematically expressed as
\begin{equation}\label{Eq-crysTwin}
\bfm{R} \bfm{U}_\text{B}-\bfm{U}_\text{A}= \bfm{a} \otimes \bfm{l},
\end{equation}
which is called the \emph{twinning equation}. In Eq.~\eqref{Eq-crysTwin}, $\bfm{U}_\text{A}$ and $\bfm{U}_\text{B}$ represent the (symmetric) transformation stretch tensors of the two variants involved (known from crystallography) and the unknowns are the twinning shear vector $\bfm{a}$, the normal to the interface $\bfm{l}$, and the rotation tensor $\bfm{R}$. In an analogous manner, in the case of an interface mediating a single variant of martensite and a twinned martensite, which is referred to as M--MM interface, the compatibility equation takes the form
\begin{equation}\label{Eq-crysM-MM}
\hat{\bfm{R}} \left( \lambda^0 \bfm{R} \bfm{U}_\text{B} +(1-\lambda^0)\bfm{U}_\text{A}\right)-\bfm{U}_\text{A}=\bfm{b} \otimes \bfm{m},
\end{equation}
where $\lambda^0$ represents the twin volume fraction and is here chosen arbitrarily in the range $0 < \lambda^0 < 1$, while the unknowns are the shear vector $\bfm{b}$, the normal to the interface $\bfm{m}$, and the rotation tensor $\hat{\bfm{R}}$. Note that the interface normals $\bfm{l}$ and $\bfm{m}$ refer to the undeformed configuration of austenite. The solution procedure for the twinning equation~\eqref{Eq-crysTwin} and the M--MM interface equation~\eqref{Eq-crysM-MM} can be found in the references cited above. To provide further clarity and to serve as an example, we present below the solution for a selected volume fraction of $\lambda^0=0.3$.

In the present paper, the focus of our main analysis is on the type-I twin in CuAlNi shape memory alloy (the case of type-II twin is commented in \ref{sec-twintype}). The martensitic transformation in this alloy proceeds via a cubic-to-orthorhombic structural change and involves six martensite variants. Among the martensite variant pairs with type-I twin relation, we have selected a representative pair $(\text{A},\text{B})=(1,3)$, which is characterized by the following transformation stretch tensors (here and below, all tensor and vector components are given in the austenite cubic basis),
\begin{equation}\label{Eq-UAUB}
\bfm{U}_\text{A}=\bfm{U}_1 =
\begin{pmatrix}
(\alpha+\gamma)/2 & 0 & (\alpha-\gamma)/2 \\
     0 &\beta& 0 \\
    (\alpha-\gamma)/2 &0& (\alpha+\gamma)/2 \\
\end{pmatrix}, \quad 
\bfm{U}_\text{B}=\bfm{U}_3 =
\begin{pmatrix}
(\alpha+\gamma)/2 & (\alpha-\gamma)/2 & 0 \\
(\alpha-\gamma)/2 &(\alpha+\gamma)/2& 0 \\
   0 &0& \beta \\
\end{pmatrix},
\end{equation}
with the stretch parameters $\alpha=1.0619$, $\beta=0.9178$ and $\gamma=1.023$ \citep{Bhattacharya2003}. The solutions of the twinning equation \eqref{Eq-crysTwin} and the M--MM interface equation \eqref{Eq-crysM-MM} for $\lambda^0=0.3$ are then obtained as follows
\begin{align}\label{Eq-sol03}
\begin{aligned}
\bfm{a}&=(-0.0515 , -0.1637 , -0.1869), \\
\bfm{l}&=(0,-0.7071,0.7071),
\end{aligned}
\quad
\begin{aligned}
\bfm{b}&=(0.0003,0.0563,-0.0535), \\
\bfm{m}&=(0.2272,0.6226,0.7486),
\end{aligned}
\end{align}
and the corresponding rotation tensors are given by
\begin{equation}\label{Eq-sol03R}
\bfm{R}=\begin{pmatrix}
\phantom{-}0.9997 & 0.0163 & -0.0185 \\
-0.0185 & 0.9918 & -0.1262 \\
\phantom{-}0.0163 & 0.1265 & \phantom{-}0.9918
\end{pmatrix}, \quad \hat{\bfm{R}}=\begin{pmatrix}
\phantom{-}0.9999 & -0.0117 & 0.0108 \\
\phantom{-}0.0108 & \phantom{-}0.9970 & 0.0765 \\
-0.0116 & -0.0763 & 0.9970
\end{pmatrix}.
\end{equation}
It should be noted that $\bfm{b}$ and $\bfm{m}$ presented in Eq.~\eqref{Eq-sol03} correspond to the non-trivial solution of the M--MM interface. A trivial solution is simply obtained as $\bfm{m}=\bfm{l}$, $\bfm{b}=\lambda^0\bfm{a}$ and $\bfm{R}=\bfm{I}$, where $\bfm{I}$ is the identity tensor, see a more detailed discussion in \citep{stupkiewicz2012almost}.

\section{Phase-field model for twinning}\label{sec-PF}
A conventional phase-field model of twinning is utilized in this {study}. In this section, we provide a concise description of the model and briefly discuss its finite-element implementation. For more details, interested readers are referred to the multi-phase versions of the model that have been developed in our previous studies \citep{rezaee2020phase,tuuma2021phase}, see also \cite{tuuma2016size} for an earlier version featuring hierarchical order parameters.

The deformation gradient $\bfm{F}$ and the non-conserved order parameter $\phi$ constitute the primary variables in the model. The finite-strain kinematic description relies on the multiplicative decomposition of the deformation gradient $\bfm{F}$ into the elastic part $\bfm{F}^\text{e}$ and the part $\bfm{F}^\text{t}$ associated with the twinning transformation, viz.,
\begin{equation}
\bfm{F}=\bfm{F}^\text{e} \bfm{F}^\text{t}, \quad \bfm{F}=\nabla \bm{\upvarphi},
\end{equation}
where $\bm{\upvarphi}$ denotes the deformation mapping from the reference configuration to the current configuration, $\bfm{x}=\bm{\upvarphi}(\bfm{X})$. Within the context of twinning, where only two martensite variants are involved, a single order parameter $\phi$ is adequate to characterize the material state. In the present model, the order parameter $\phi$ is interpreted as the relative twin volume fraction and is bounded within the range $0 \leq \phi \leq 1$, where $\phi=0$ and $\phi=1$ correspond to pure martensite variants (here, variant A and variant B, respectively), while the intermediate values $0 < \phi <1$ represent the diffuse twin interfaces.

Among the available formulations for the transformation deformation gradient $\bfm{F}^\text{t}$, the rank-one mixing rule is adopted \citep{levitas2009displacive,clayton2011phase}
\begin{equation}\label{Eq-rankone}
\bfm{F}^\text{t}=\bfm{U}_\text{A}+ \phi \, \bfm{a} \otimes \bfm{l},
\end{equation}
{which is defined explicitly in terms of one of the solutions ($\bfm{a},\bfm{l}$) of the twinning equation~\eqref{Eq-crysTwin} (recall that the twinning equation has two solutions). By construction, the transformation deformation gradient $\bfm{F}^\text{t}$ in Eq.~\eqref{Eq-rankone} is rank-one connected to $\bfm{U}_\text{A}$ for any value of $0 \leq \phi \leq 1$, so that a planar diffuse interface with the normal $\bfm{l}$ is fully compatible. Note, however, that compatibility is not ensured \emph{within} a diffuse interface (i.e., for $0 < \phi < 1$) that has the orientation of the other solution of the twinning equation, and elastic strains are then needed to accommodate the incompatibility within such an interface, see the related discussion in Remark 2.5 in \citep{rezaee2022deformation}.}

At this point, a variational formulation of the model is derived following the approach of Hildebrand and Miehe \citep{hildebrand2012phase}, see also \citep{tuuma2016size}. This implies that the model unknowns ($\bm{\upvarphi}, \phi$), and therefore the microstructure evolution, are governed by the minimization problem formulated for the total incremental potential of the system,
\begin{equation}
\Pi = \Delta \mathcal{F}+\mathcal{D}_\tau \; \rightarrow \; \min_{\substack{\bm{\upvarphi},\phi}}
\end{equation}
which is subject to the inequality constraint for the order parameter $0 \leq \phi \leq 1$. Here, {$\Delta \mathcal{F}$} denotes {the increment of the total} Helmholtz free energy, and $\mathcal{D}_\tau$ denotes the {incremental} dissipation potential. We consider the model to be constrained to isothermal processes. Consequently, the total Helmholtz free energy $\mathcal{F}$ and the dissipation potential $\mathcal{D}_\tau$ are defined (for the entire body $B$) as
\begin{equation}
\mathcal{F}=\int_B \left( F_\text{el} + F_\text{int} \right) \rd V, \quad \mathcal{D}_\tau=\int_B D_\tau \, \rd V.
\end{equation}
It thus remains to define the Helmholtz free energy contributions, namely the elastic strain energy $F_\text{el}$ and the interfacial energy $F_\text{int}$, and also the (time-discrete) dissipation function $D_\tau$. Note that the absence of a chemical energy contribution in the Helmholtz free energy is justified by the assumption that, under stress-free conditions, the martensite variants are energetically equivalent.

Following our previous works \citep{rezaee2020phase,tuuma2021phase}, a Hencky-type anisotropic elastic strain energy is considered, which takes the form
\begin{equation}\label{Eq-elasticEn}
F_\text{el}=\frac{1}{2} (\det \bfm{F}^\text{t}) \bfm{H}^\text{e} \cdot \mathbb{L} \bfm{H}^\text{e}, \quad \bfm{H}^\text{e}=\frac{1}{2} \log \bfm{C}^\text{e}, \quad \bfm{C}^\text{e}=(\bfm{F}^\text{e})^\text{T}\bfm{F}^\text{e},
\end{equation}
where $\bfm{H}^\text{e}$ is the logarithmic elastic strain, $\bfm{C}^\text{e}$ is the elastic right Cauchy-Green tensor and $\mathbb{L}=(1-\phi)\mathbb{L}_\text{A}+\phi \mathbb{L}_\text{B}$ is the (effective) fourth-order elastic stiffness tensor, obtained by Voigt-type averaging of the elastic stiffness tensors of martensite variants A and B, see \citep{CowinMehrabadi1995} for the general form of an elastic stiffness tensor with orthorhombic symmetry.

On the other hand, a double-obstacle potential with an isotropic gradient energy term is adopted for the interfacial energy \citep{steinbach2009phase},
\begin{equation}\label{Eq-intEn}
F_\text{int}=\frac{4\gamma_\text{tw}}{\pi \ell} \left( \phi (1-\phi) + \ell^2 \nabla \phi \cdot \nabla \phi \right),
\end{equation}
where $\gamma_\text{tw}$ is the interfacial energy {density} (per unit area) associated with the martensite--martensite (twin) interface and $\ell$ is the corresponding interface thickness parameter. The interfacial energy of the form~\eqref{Eq-intEn} leads to a theoretical (i.e., under stress-free conditions) interface thickness of $\pi \ell$.

The final component of the model to be specified is the time-discrete dissipation function $D_\tau$. In line with the conventional phase-field modeling, a viscous dissipation is employed here, which is expressed as
\begin{equation}\label{Eq-diss}
D_\tau=\tau D=\frac{\tau}{2m} \left(\frac{\phi-\phi_n}{\tau}\right)^2,
\end{equation}
where $m$ is the interface mobility parameter, $\tau$ is the time increment and $\phi_n$ is the order parameter at the previous time step. It is noteworthy that the form~\eqref{Eq-diss} is obtained by applying the backward-Euler method to integrate the rate-potential $D=(1/2m) \dot{\phi}^2$ {which is expressed in terms of $\dot{\phi}$, the rate of the order parameter}.

We now briefly outline the most important aspects of the finite-element implementation of the phase-field model described above. The actual unknowns of the model in the implementation are the displacement field $\bfm{u}=\bm{\upvarphi}-\bfm{X}$ and the order parameter $\phi$. As will be discussed in Section~\ref{sec-res}, our analysis is restricted to generalized plane strain condition. Thus, spatial discretization is done by using isoparametric 8-noded serendipity elements (with reduced $2 \times 2$ Gauss integration rule) for the displacement field $\bfm{u}$ and 4-noded bilinear elements for the order parameter $\phi$. The resulting discretized nonlinear equations are solved in a monolithic fashion by using the Newton method. The penalty regularization method is employed to enforce the inequality constraint for the order parameter, $0 \leq \phi \leq 1$, as done in our prior studies involving multiple order parameters \citep{rezaee2020phase,tuuma2021phase}. 

For an efficient and reliable computer implementation, the AceGen system is used \citep{korelc2009automation,korelc2016}, which features automatic differentiation and code simplification capabilities, and thereby, guarantees an exact computation of the tangent matrix. The simulations are carried out by using AceFEM, a finite-element environment closely connected with AceGen.

\section{Phase-field simulation results}\label{sec-res}

In this section, we present and discuss the results obtained from our phase-field simulations. The setup of the problem and the material parameters are outlined in Section~\ref{sec-setup}, while the quantitative measures which are used for analyzing the simulation results are described in Section~\ref{sec-eqs}. The discussion of the simulation results commences with the analysis of a representative case in Section~\ref{sec-rep}. Subsequently, the effect of twin spacing and the related size effects are studied in Section~\ref{sec-size}. Finally, in Section~\ref{sec-vol}, the effect of twin volume fraction is investigated. 

\subsection{Problem setup and material parameters}\label{sec-setup}
\begin{figure}
\centering
\hspace*{-1.8cm}
\includegraphics[width=1.2\textwidth]{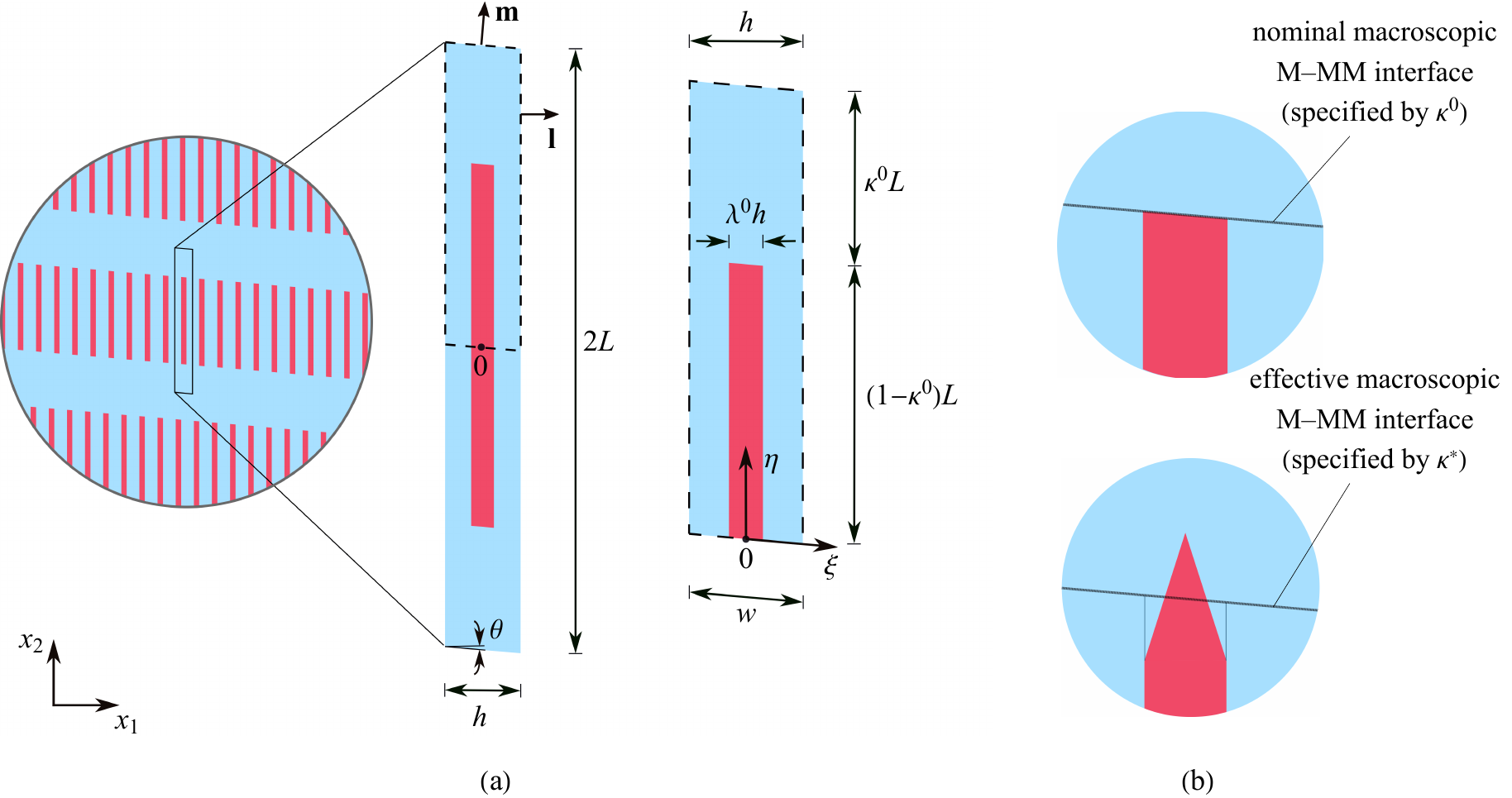}
\caption{(a) The setup of the problem and the initial conditions, and (b) a schematic illustration of the nominal and effective (energy-minimizing) macroscopic M--MM interfaces. In panel (a), the sketch in the middle depicts the full periodic domain and the sketch on the right depicts the actual computational domain used in the simulations. To enhance the clarity of the sketches, the aspect ratio of the objects has been considerably decreased beyond their actual proportions.}
\label{fig-setup}
\end{figure}

The purpose of our computational study is to conduct a detailed analysis of the macroscopic M--MM interface in a CuAlNi single crystal (see Fig.~\ref{fig-seiner}) by using the phase-field model presented in Section~\ref{sec-PF}. Within such a macroscopic interface, a microstructured transition layer (of some finite width) is formed in which the local incompatibility between the (macroscopically) homogeneous phases of single martensite variant and twinned martensite is accommodated by elastic strains. It is assumed in the present study that this transition layer is morphologically periodic along the M--MM interface with the period being the twin spacing $h$ containing a twin pair. Meanwhile, outside of the transition layer and far away from the M--MM interface, the elastic micro-strain energy and the related stresses are expected to tend to zero. Accordingly, in the finite-element simulations, the domain under study is chosen to be sufficiently long in the direction parallel to the twin interfaces, i.e., a geometrical aspect ratio of at least $2L/h$=40 is selected where $2L$ denotes the height of the domain, and with the periodic boundary conditions enforced at the corresponding edges, see Fig.~\ref{fig-setup}(a). It is noteworthy that our computational problem is closely related to that of T\r{u}ma and Stupkiewicz \citep{tuuma2016phase} on the austenite--twinned martensite interfaces, see also \citep{maciejewski2005elastic} for the related sharp-interface modeling study. 

Nevertheless, it turned out from our simulations that the far-field elastic micro-strain energy is non-zero, indicating that the elastic strains (and hence energy) are not confined to a transition layer along the M--MM interface. Thereby, as will be elaborated in Sections~\ref{sec-eqs} and \ref{sec-rep}, corrections ought to be made in order to subtract the far-field energy contributions from the energy of the macroscopic interface, as accomplished in \citep{petryk2010interfacial}. It should be pointed out that the far-field energy contributions diminish by increasing the geometrical aspect ratio $2L/h$. However, acquiring zero far-field energy contributions, as confirmed by our auxiliary simulations, would require a very long computational domain, which, due to the computational restrictions, is not feasible.

As discussed in Section~\ref{sec-crys}, the twinning plane normal $\bfm{l}$ and the M--MM interface normal $\bfm{m}$ can be obtained from the crystallographic theory. Accordingly, the domain under study is oriented such that the problem refers to a plane that contains both $\bfm{l}$ and $\bfm{m}$, as shown in Fig.~\ref{fig-setup}(a). This means that the global $x_1$ axis aligns with $\bfm{l}$ and the global $x_2$ axis deviates from $\bfm{m}$ by a characteristic angle $\theta$. As a result, the domain takes the shape of a parallelogram, and the angle $\theta$ measures the deviation of the parallelogram from a rectangular shape. A generalized plane strain condition is considered in the simulations, which implies that, while the problem is independent of the out-of-plane spatial dimension, it accounts for a non-zero out-of-plane displacement, e.g.,\ \citep{maciejewski2005elastic,tuuma2016phase}. 

A deformation-controlled loading is applied through prescribing a constant average (overall) deformation gradient $\bar{\bfm{F}}$. Subsequently, the microstructure is allowed to attain a steady (equilibrium) state (i.e., the minimum of the total free energy is obtained through a viscous evolution of the microstructure). This state is then taken as the subject of the analysis. The average deformation gradient $\bar{\bfm{F}}$ is defined in the following form
 \begin{equation}\label{Eq-avedef}
   \bar{\bfm{F}}=\kappa^0 \bfm{F}_1+(1-\kappa^0)(\lambda^0 \bfm{F}_2+(1-\lambda^0)\bfm{F}_3),
\end{equation}
with the individual deformation gradients $\bfm{F}_1$, $\bfm{F}_2$ and $\bfm{F}_3$ being equal to
\begin{equation}
\bfm{F}_1=\bfm{U}_\text{A}, \quad \bfm{F}_2=\hat{\bfm{R}} \bfm{R} \bfm{U}_\text{B}, \quad \bfm{F}_3=\hat{\bfm{R}} \bfm{U}_\text{A}.
\end{equation}
Here, $\bfm{U}_\text{A}=\bfm{U}_1$ and $\bfm{U}_\text{B}=\bfm{U}_3$, see Eq.~\eqref{Eq-UAUB}, represent the transformation stretch tensors of the two martensite variants involved, and the rotations $\hat{\bfm{R}}$ and $\bfm{R}$ come from the crystallographic theory, see Eq.~\eqref{Eq-sol03R}. Accordingly, the average deformation gradient $\bar{\bfm{F}}$ corresponds to (theoretically) stress-free conditions. Eq.~\eqref{Eq-avedef} involves two volume fractions, namely $\lambda^0$ and $\kappa^0$ (referred to as `nominal' volume fractions in the sequel). The former controls the relative twin volume fraction and is an input in the crystallographic theory equations, cf.~Eq.~\eqref{Eq-crysM-MM}, while the latter controls the overall volume fraction of the single martensite {and twinned martensite regions} within the computational domain and is adjusted such that the domain of twinned martensite is sufficiently large to accommodate a rather long needle-shaped microstructure. Note that the initial state of the system is set by prescribing the order parameter $\phi$ in accordance with the nominal volume fractions $\lambda^0$ and $\kappa^0$, see Fig.~\ref{fig-setup}(a).

Our simulations revealed negligible discrepancies between the results of type-I and type-II twins. As a consequence, our primary focus in this study is on the analysis of type-I twins, while the striking resemblance between the simulation results of type-I and type-II twins is illustrated in \mbox{\ref{sec-twintype}}. {As such, the mixing rule~\eqref{Eq-rankone} is formulated in terms of the type-I solution of the twinning equation, as given in Eq.~\eqref{Eq-sol03}.}

The computational domain is discretized by using a uniform finite-element mesh (unless stated otherwise). The size of the elements $d$ is set such that the mesh is fine enough to properly resolve the interfaces and to capture the subtle features of the resulting microstructure. More specifically, a ratio of approximately 5 is considered between $\pi \ell$ and $d$, where the former is the theoretical interface thickness. Periodic boundary conditions are enforced on both the displacement field $\bm{u}$ and the order parameter $\phi$. To reduce the computational cost, the two-fold rotational symmetry of the microstructure about the central point (see point 0 in Fig.\ref{fig-setup}(a)) is exploited, and thereby, only half of the domain (of the size $h \times L$) is computed. Accordingly, the anti-periodicity of the displacement field $\bm{u}$ and the symmetry of the order parameter $\phi$ with respect to the point 0 are enforced at the bottom edge, and similarly at the top edge.

The following material parameters are used in all the simulations. The anisotropic elastic constants of orthorhombic martensite, namely ${c}_{11}=189$, ${c}_{22}=141$, ${c}_{33}=205$, ${c}_{44}=54.9$, ${c}_{55}=19.7$, ${c}_{66}=62.6$, ${c}_{12}=124$, ${c}_{13}=45.5$, $ {c}_{23}=115$ (all in GPa), are adopted from the available literature data \citep{suezawa1976behaviour,yasunaga1983measurement}. The interfacial energy density is selected as $\gamma = 0.02\,\mathrm{J/m^2}$, see e.g., \citep{tuuma2016phase}. Finally, the mobility parameter $m$ takes the value of $m=1$ (MPa s)$^{-1}$. Note that our analysis is limited to the steady-state microstructure, and not its evolution process. Therefore, the mobility parameter $m$ acts merely as a regularization parameter and its value does not affect the final results.

\subsection{Quantitative description of the microstructure}\label{sec-eqs}
Throughout the analysis, in addition to the examination of morphological features of the predicted microstructures, we employ a set of quantitative measures to characterize the microstructures and compare them across different cases. The selected measures are established based on the following averaging operations,
\begin{equation}\label{Eq-average}
\langle \cdot \rangle=\left.\!\langle \cdot \rangle\right|_\eta= \frac{1}{w} \int_{-w/2}^{w/2} \left( \cdot \right)  \, \rd \xi, \quad \{ \cdot \}=\frac{1}{L}\int_0^{L}\left.\!\langle \cdot \rangle\right|_\eta \rd \eta=\frac{1}{w L} \int_0^{L} \left( \int_{-w/2}^{w/2} \left( \cdot \right)  \, \rd \xi \right) \rd \eta,
\end{equation}
see Fig.~\ref{fig-setup}(a) for the definition of $\xi$ and $\eta$ coordinates. Accordingly, through the width-averaging operation $\langle \cdot \rangle$, the average order parameter $\langle \phi \rangle$, the integrated elastic strain energy $h \langle F_\text{el} \rangle$, cf.~Eq.~\eqref{Eq-elasticEn}, and the integrated interfacial energy $h \langle F_\text{int} \rangle$, cf.~Eq.~\eqref{Eq-intEn}, are obtained. Note that these averages can be evaluated at arbitrary height, thus $\langle \cdot \rangle=\left.\!\langle \cdot \rangle\right|_\eta$. At the same time, the respective overall quantities are determined via the volume-averaging operation $\{ \cdot \}$, namely the overall order parameter $\{ \phi \}$, the total elastic strain energy $\mathcal{F}_\text{el}=hL \{ F_\text{el}\}$ and the total interfacial energy $\mathcal{F}_\text{int}=hL \{ F_\text{int}\}$.

In order to effectively characterize the overall elastic micro-strain energy of the macroscopic M--MM interface, we define the energy-based measures $\gamma_\text{el}^\text{tot}$ and $\Gamma_\text{el}^\text{tot}$ as
\begin{equation}\label{Eq-microstrain}
\gamma_\text{el}^\text{tot}=\frac{\mathcal{F}_\text{el}}{w},\quad \Gamma_\text{el}^\text{tot}=\frac{\gamma_\text{el}^\text{tot}}{h},
\end{equation}
where $\gamma_\text{el}^\text{tot}$ represents the elastic micro-strain energy per unit area of the M--MM interface, while the energy factor $\Gamma_\text{el}$ measures the dependence of the energy on the microstructure. Both $\gamma_\text{el}^\text{tot}$ and $\Gamma_\text{el}^\text{tot}$ provide a quantification of the elastic strain energy of the M--MM interface, however, the energy factor $\Gamma_\text{el}^\text{tot}$ is of particular importance, as it filters out the first-order dependence of $\gamma_\text{el}^\text{tot}$ on the twin spacing $h$, and thus in this sense it can be considered size-independent \citep{maciejewski2005elastic,tuuma2016phase}.

The results of our simulations reveal the presence of non-zero values of the elastic strain energy $h \langle F_\text{el} \rangle$ far from the M--MM interface, i.e., at the upper and lower boundaries of the computational domain at $\eta=0$ and $\eta=L$, see, for instance, Fig.~\ref{Fig-repProfiles} in Section~\ref{sec-rep} and the associated discussion. It is therefore desirable to mitigate these far-field energy contributions by correcting the elastic micro-strain energy measures. To this end, we first define the effective volume fractions $\lambda^\ast$ and $\kappa^\ast$ as follows
\begin{equation}\label{Eq-effective} 
\left.\!\lambda^\ast=\langle \phi \rangle \right|_{\eta=0}, \quad \kappa^\ast=1-\frac{\{ \phi \}}{\lambda^\ast}.
\end{equation}
Note that while $\kappa^0$, cf.~Eq.~\eqref{Eq-avedef}, specifies the nominal position of the M--MM interface, $\kappa^*$ specifies the corresponding effective (actual) position, as delineated in Fig.~\ref{fig-setup}(b). Next, the `corrected' total elastic strain energy $\mathcal{F}_\text{el}^\text{corr}$ is calculated upon subtracting the contributions of the far-field energies $\mathcal{F}_\text{el}^{\infty,0}$ and $\mathcal{F}_\text{el}^{\infty,L}$ from the total elastic strain energy $\mathcal{F}_\text{el}$, see \citep{petryk2010interfacial}, viz.,
\begin{equation}\label{Eq-adj}
\mathcal{F}_\text{el}^\text{corr}=\mathcal{F}_\text{el}-\mathcal{F}_\text{el}^{\infty,0}-\mathcal{F}_\text{el}^{\infty,L},
\end{equation}
where
\begin{equation}\label{Eq-far}
\mathcal{F}_\text{el}^{\infty,0}=(1-\kappa^\ast) h L \langle F_\text{el} \rangle|_{\eta=0}, \quad \mathcal{F}_\text{el}^{\infty,L}=\kappa^\ast h L \langle F_\text{el} \rangle|_{\eta=L}.
\end{equation}
Note that the far-field energies $\mathcal{F}_\text{el}^{\infty,0}$ and $\mathcal{F}_\text{el}^{\infty,L}$ can be also computed by using the nominal volume fraction $\kappa^0$ (by simply substituting $\kappa^*$ by $\kappa^0$). This aspect is discussed in the subsequent sections. 
Finally, the new energy-based measures $\gamma_\text{el}$ and $\Gamma_\text{el}$ are defined as
\begin{equation}\label{Eq-corrGamma}
\gamma_\text{el}=\frac{\mathcal{F}_\text{el}^\text{corr}}{w}, \quad \Gamma_\text{el}=\frac{\gamma_\text{el}}{h}.
\end{equation}

An additional energy-based measure that is consistently used to assess the macroscopic M--MM interface is the excess interfacial energy density $\gamma_\text{int}^\text{exs}$ defined as
\begin{equation}\label{Eq-excess}
\gamma_\text{int}^\text{exs}=\frac{\mathcal{F}_\text{int}^\text{exs}}{w}, \quad \mathcal{F}_\text{int}^\text{exs}=\mathcal{F}_\text{int}-\mathcal{F}_\text{int}^\text{ref}.
\end{equation}
Here, $\mathcal{F}_\text{int}^\text{ref}$ represents the total interfacial energy related to the nominal (needle-less) M--MM interface (i.e., when only the two parallel planar twin interfaces are accounted for) and is calculated either based on the effective volume fraction $\kappa^\ast$, i.e., $\mathcal{F}_\text{int}^\text{ref}=2(1-\kappa^\ast)L \gamma_\text{tw}$, or based on the nominal volume fraction $\kappa^0$, i.e., $\mathcal{F}_\text{int}^\text{ref}=2(1-\kappa^0)L \gamma_\text{tw}$. Recall that $\gamma_\text{tw}$ is the interfacial energy density associated with the local martensite--martensite interface, see Eq.~\eqref{Eq-intEn}. In fact, $\gamma_\text{int}^\text{exs}$ in Eq.~\eqref{Eq-excess} represents the extra interfacial energy density resulting from the difference between the predicted microstructure and the needle-less microstructure, see the schematic representation of the respective M--MM interfaces in Fig.~\ref{fig-setup}(b).

\subsection{Modeling M--MM interface: a representative study}\label{sec-rep}
In this section, we present the analysis of a representative study, with the aim to elucidate the individual characteristics of the simulated microstructures, as a preliminary step prior to examining their collective macroscopic responses. The computational domain considered in this study has dimensions of $h \times L=70 \times 1400$ nm$^2$, the selected nominal volume fractions are $\kappa^0=0.4$ and $\lambda^0=0.3$, and the interface thickness parameter is adopted as $\ell=1$ nm. The computational domain is discretized into approximately 250\,000 elements of the size $d=0.625$ nm, thus resulting in approximately 2.5 million degrees of freedom. 

\begin{figure}
\centering
\hspace*{-2.5cm}
\includegraphics[width=1\textwidth]{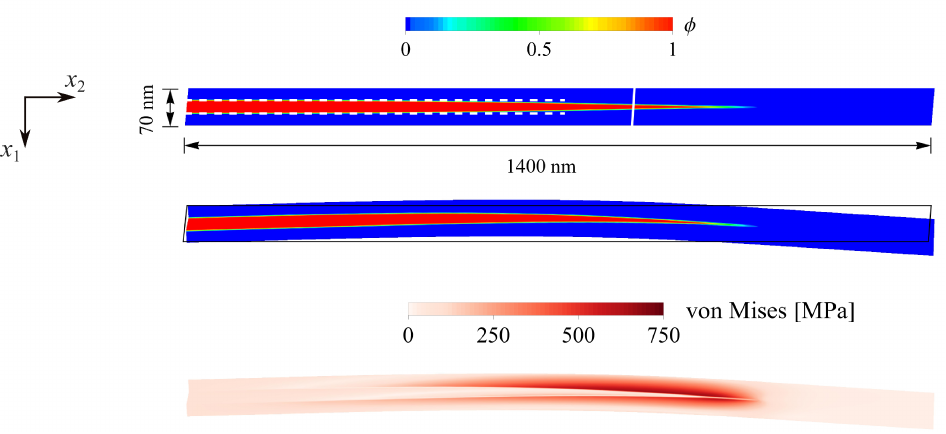}
\caption{Simulation results corresponding to the representative study: the steady-state microstructure containing a needle-shaped domain of variant B in both the undeformed and deformed configurations. The color contours represent the fields of the order parameter $\phi$ and von Mises stress. The parallel dashed lines overlaid on the undeformed microstructure delineate the trajectory of the planar interfaces extended up to the effective position of the M--MM interface, determined based on $\kappa^*$, while the inclined solid line indicates the nominal position of the M--MM interface, determined based on $\kappa^0$. {The place at which the dashed lines and the actual interfaces begin to diverge} indicates the onset of the wedge-shaped region of the microstructure.}
\label{Fig-repMic}
\end{figure}

Fig.~\ref{Fig-repMic} depicts the simulation results in terms of the steady-state microstructure. The microstructure is visualized in both the undeformed and deformed configurations, and is represented by the spatial distribution of the order parameter $\phi$ and of the von Mises stress. As shown in Fig.~\ref{Fig-repMic}, the microstructure in its steady state has developed a distinctive needle-shaped domain of martensite variant B. While the needle appears to be straight in the undeformed configuration, it exhibits a visible bending in the deformed configuration with a bending angle of approximately 5$^\circ$ (measured at the needle apex) with respect to the longitudinal axis (global $x_2$ axis). The observed bending of the needle conforms with the experimental observations, e.g., \citep{boullay2001bending,seiner2011finite}, as well as with the previous modeling analyses, e.g., {\citep{li1999theory,finel2010phase,seiner2011finite,conti2020geometry}}. Another noteworthy feature of the microstructure pertains to the excessive diffuseness of the needle apex. Our auxiliary simulations, {aimed at investigating the effect of the interface thickness parameter $\ell$ on the microstructure}, confirmed that such excessive diffuseness does not represent a physical characteristic of the microstructure, but rather a numerical artifact arising from the phase-field modeling framework. {It was observed that reducing $\ell$} results in a significantly less diffuse needle apex. {Nevertheless, it is important to note that a smaller $\ell$} requires a relatively finer finite-element mesh, {which renders the computations} excessively costly, and thus has not been considered in our main simulations.

From the distribution of the von Mises stress, we observe that the stress is predominantly concentrated in the areas surrounding the curved interfaces (within variant A) close to the needle apex. Conversely, within the needle itself (within variant B), the stress is considerably lower. Interestingly, our von Mises stress distribution shows qualitative (and to some extent quantitative) similarities with the stress distribution obtained by Seiner et al.\ \citep{seiner2011finite}, in particular, the stress distribution related to the `optimal' case that results from the minimization of elastic energy, see Fig.~8 therein.

\begin{figure}
\centering
\hspace*{-1.2cm}
\includegraphics[width=1.1\textwidth]{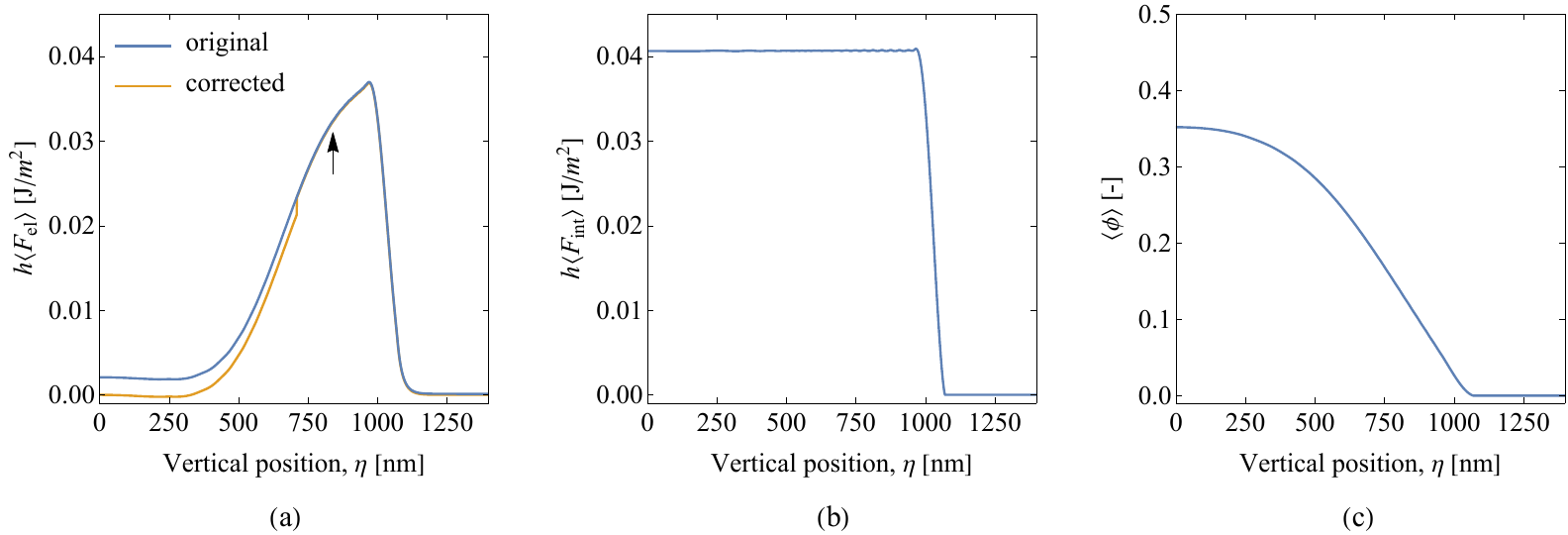}
\caption{Simulation results corresponding to the representative study: the longitudinal profiles of (a) the integrated elastic strain energy $h \langle F_\text{el} \rangle$, (b) the integrated interfacial energy $h \langle F_\text{int} \rangle$, and (c) the average order parameter $\langle \phi \rangle$. The yellowish curve in panel (a) refers to the case where the profile of the elastic strain energy is corrected by means of subtracting the far-field energy contributions. The arrow in panel (a) indicates the position where the division of the curve into two segments (see the text) would take place if the nominal volume fraction $\kappa^0$ were used for the correction.}
\label{Fig-repProfiles}
\end{figure}

We continue the discussion by examining the longitudinal profiles of the integrated elastic strain energy $h \langle f_\text{el} \rangle$, integrated interfacial energy $h \langle f_\text{int} \rangle$ and the average order parameter $\langle \phi \rangle$, see Fig.~\ref{Fig-repProfiles}. It is immediate to see that the elastic strain energy reaches its peak within the region occupied by the needle apex and then decays rapidly as it moves away from this region. Contrary to the notion that the elastic micro-strain energy vanishes far away from the M--MM interface, we observe that the far-field elastic strain energy is non-zero and is equal to $\left.\! h \langle F_\text{el}\rangle \right|_{\eta=0}=2.1 \times 10^{-3}$ J/m$^2$ and $\left.\! h \langle F_\text{el} \rangle \right|_{\eta=L}=1.2 \times 10^{-4}$ J/m$^2$. Notably, the presence of non-zero energy contribution at $\eta=0$ is directly linked to the observed noticeably higher value of the effective volume fraction $\lambda^*=0.35$ compared to the nominal volume fraction $\lambda^0=0.3$. And this, in turn, leads to a deviation between the effective volume fraction $\kappa^*=0.49$ and the nominal volume fraction $\kappa^0=0.4$, as highlighted by the dashed and solid white lines overlaid on the undeformed microstructure in Fig.~\ref{Fig-repMic}. The underlying cause of this omnipresent discrepancy can be sought in the dominant role of the interfacial energy, stemming from the limited size of the computational domain, and will be further discussed in Section~\ref{sec-size}. 

As discussed previously in Sections~\ref{sec-setup} and \ref{sec-eqs}, we opt to mitigate the far-field energy contributions by introducing the corrected elastic micro-strain energy measures $\gamma_\text{el}$ and $\Gamma_\text{el}$, see Eq.~\eqref{Eq-corrGamma}. As an illustration of this correction, the corrected profile of the integrated elastic strain energy $h \langle F_\text{el} \rangle$ is depicted in Fig.~\ref{Fig-repProfiles}(a). This correction is accomplished by dividing the curve into two segments using the effective volume fraction $\kappa^\ast$, where the first segment spans from $\eta=0$ to $\eta=(1-\kappa^\ast)L$, and the second segment spans from $\eta=(1-\kappa^\ast)L$ to $\eta=L$. Subsequently, we subtract the far-field elastic strain energy $\left.\! h \langle F_\text{el}\rangle \right|_{\eta=0}=2.1 \times 10^{-3}$ J/m$^2$ from the first segment and $\left.\! h \langle F_\text{el} \rangle \right|_{\eta=L}=1.2 \times 10^{-4}$ J/m$^2$ from the second segment. Note that the area beneath this corrected curve is equal to $\mathcal{F}_\text{el}^\text{corr}$, cf.~Eq.~\eqref{Eq-adj}. Alternatively, this adjustment can be also made by employing the nominal volume fraction $\kappa^0$. The division of the curve into two segments would then take place at a different position, as shown by the arrow in Fig.~\ref{Fig-repProfiles}(a).

Our discussion in this section concludes by drawing attention to the trend of the interfacial energy $h \langle F_\text{int} \rangle$. Specifically, $h \langle F_\text{int} \rangle=0.0406$ J/m$^2$ remains almost constant throughout the entire length of the twinned martensite domain. Considering that two martensite--martensite interfaces are present, the resulting interfacial energy density amounts to $h \langle F_\text{int} \rangle/2=0.0203$ J/m$^2$, which is only marginally higher than the interfacial energy density $\gamma_\text{tw}=0.02$ J/m$^2$ used in the simulations. This discrepancy is likely due to the finite resolution of the interfaces in our simulation, leading to inexact integration of the interfacial energy.

\subsection{Size effects}\label{sec-size}
This section aims to investigate the impact of twin spacing $h$ on the microstructure and the energy-based characteristics of the M--MM interface, and thereby, elucidate the related size effects. A series of simulations are carried out for twin spacing $h$ ranging from $h=20$~nm to $h=160$~nm. Note that the geometrical aspect ratio is the same in all cases, $L/h=20$. As discussed in Section~\ref{sec-setup}, in order to maintain a reasonable resolution of the predicted microstructures, the ratio of $\pi \ell/d=5$ (recall that $\pi \ell$ refers to the theoretical interface thickness and $d$ to the element size) is kept constant throughout all simulations. As such, it is not computationally feasible to perform all the simulations using a fixed interface thickness parameter $\ell$. Instead, a common strategy, as adopted in previous studies \citep{tuuma2016phase,rezaee2020phase}, is utilized in which $\ell$ and $d$ are proportionally increased, and this facilitates the extension of our analysis over a broader range of twin spacing. As it will be shown, the choice of $\ell$ has a small influence on the simulation results, confirming the validity of our analysis outcomes.

Prior to a quantitative examination of the results, it should be pointed out that within the realm of phase-field modeling, the size-dependence of the microstructure stems from the inherent length-scale of the interfacial energy and manifests itself as a result of the competition between the elastic strain energy and interfacial energy. Indeed, this fundamental premise underpins all the size-dependent characteristics observed in this study. Specifically, as the twin spacing $h$ is increased, it leads to a shift in the balance of energy from the interfacial energy to elastic strain energy, and hence the minimization of the total energy gives rise to needle-shaped microstructures with relatively longer wedges. On the contrary, for small $h$, since the interfacial energy is dominant, it is energetically favorable for the microstructure to develop a smaller area of interfaces, i.e., a relatively short domain of needle-shaped martensite. This is, however, achieved at the cost of an increase in the elastic strain energy. In particular, since the total volume fraction of variant B (quantified by $\{\phi\}$) is indirectly constrained by the displacement boundary conditions to be close to the nominal one (i.e., $\{\phi\}=(1-\kappa^*)\lambda^* \approx (1-\kappa^0)\lambda^0$), the shortening of the needle results in an increase in the twin volume fraction, hence $\lambda^* > \lambda^0$. This is then accommodated by the elastic strain energy that does not vanish far from the M--MM interface.

\begin{figure}
\centering
\hspace{-1cm}
\includegraphics[width=0.86\textwidth]{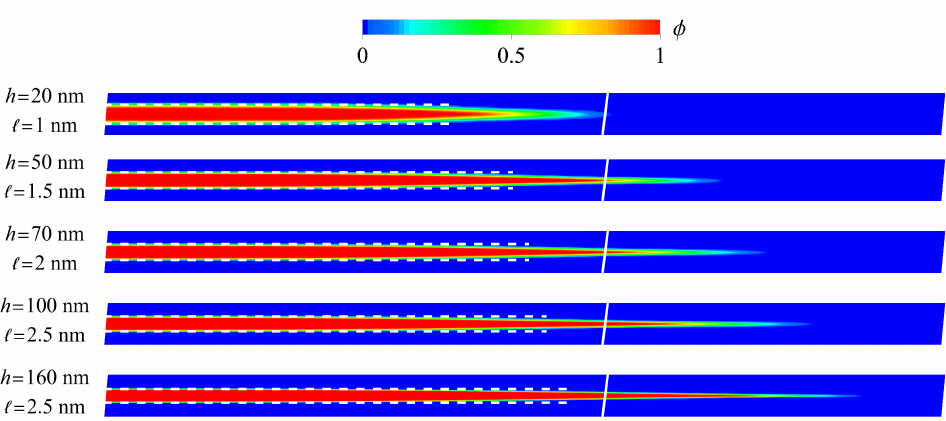}
\caption{The steady-state microstructure (represented in the undeformed configuration) for different twin spacing~$h$. The parallel dashed lines indicate the trajectory of the planar interfaces and are extended up to the effective position of the M--MM interface (specified by $\kappa^*$), while the inclined solid lines indicate the position of the nominal M--MM interface (specified by $\kappa^0$).}
\label{Fig-twinMic}
\end{figure}

\begin{figure}
\centering
\hspace{-1cm}
\includegraphics[width=0.852\textwidth]{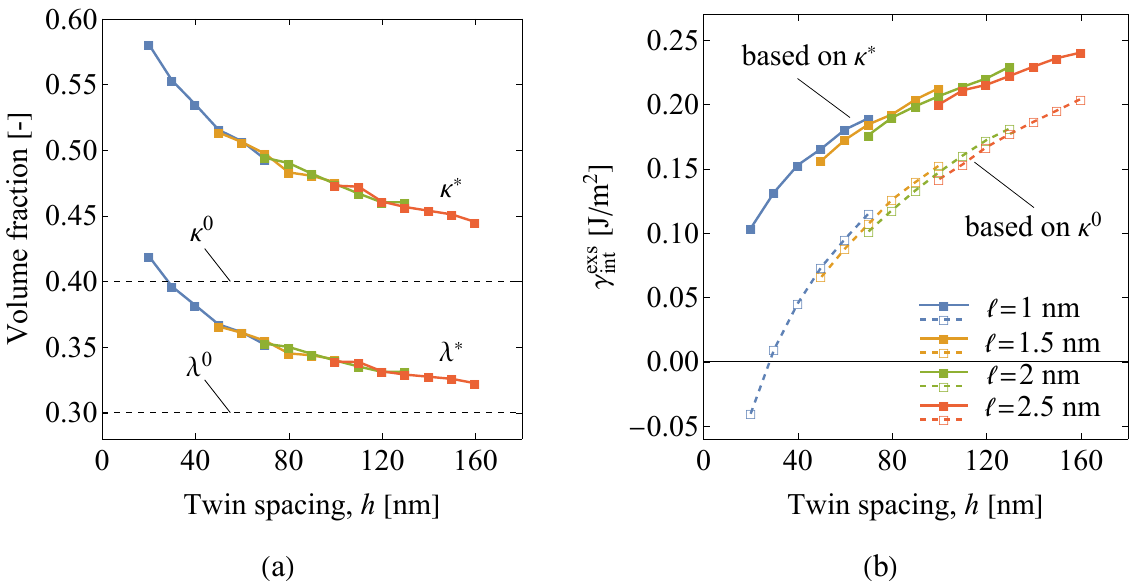}
\caption{The impact of twin spacing $h$ on (a) the effective volume fractions $\lambda^\ast$ and $\kappa^\ast$, and on (b) the excess interfacial energy density $\gamma_\text{int}^\text{exs}$. The solid and dashed curves in panel (b) correspond to cases where $\gamma_\text{int}^\text{exs}$ is determined based on the effective volume fraction $\kappa^\ast$ and nominal volume fraction $\kappa^0$, respectively, cf.~Eq.~\eqref{Eq-excess}.}
\label{Fig-excessSize}
\end{figure}

Fig.~\ref{Fig-twinMic} depicts the steady-state microstructures for different twin spacing $h$, allowing for a clear observation of the size-dependent microstructural changes described above, in particular, concerning the discrepancy between the effective and nominal volume fractions. A quantitative examination of the microstructures (see Fig.~\ref{Fig-excessSize}(a)) reveals that this discrepancy is, as expected, more pronounced for smaller $h$. As $h$ is increased, both $\lambda^\ast$ and $\kappa^\ast$ gradually tend towards their corresponding nominal values. A direct outcome of the microstructural changes depicted and quantified in Figs.~\ref{Fig-twinMic} and \ref{Fig-excessSize}(a) is reflected in the plot of excess interfacial energy density $\gamma_\text{int}^\text{exs}$ (Fig.~\ref{Fig-excessSize}(b)), which is calculated once based on the nominal volume fraction $\kappa^0$ and once based on the effective volume fraction $\kappa^\ast$, see Eq.~\eqref{Eq-excess} and the related discussion. It follows that both graphs in Fig.~\ref{Fig-excessSize}(b) exhibit similar monotonically increasing trend and that they are visibly distant, notably for smaller $h$. The monotonically increasing trend of $\gamma_\text{int}^\text{exs}$ displays the elongation of the wedge-shaped region of the microstructure as $h$ increases, and thus the more interfacial energy associated with it. 

\begin{figure}
\centering
\includegraphics[width=0.8\textwidth]{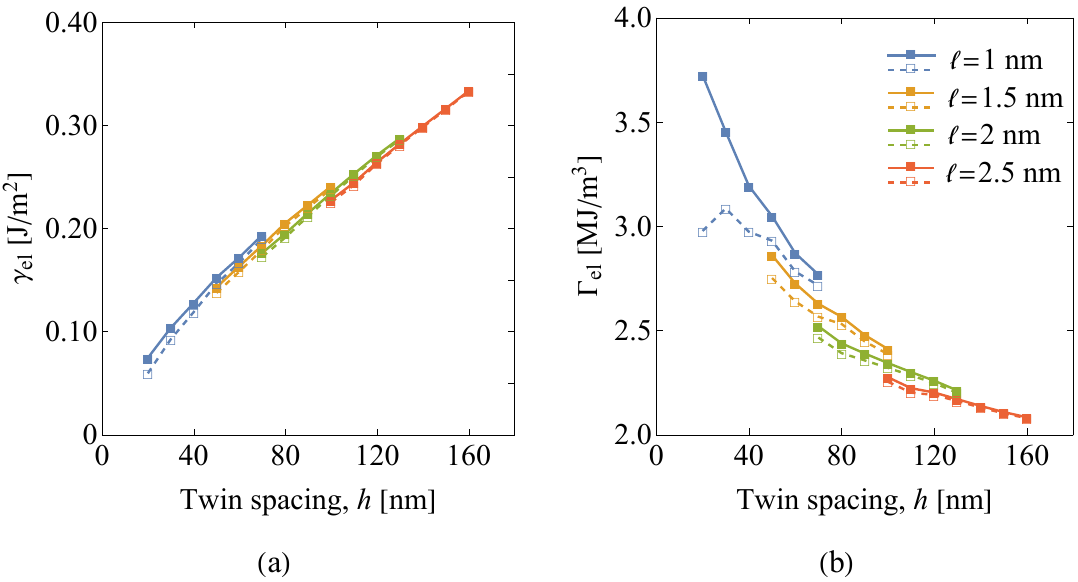}
\caption{The graphs of the elastic micro-strain energy $\gamma_\text{el}$ (a) and the corresponding energy factor $\Gamma_\text{el}$ (b) as a function of twin spacing $h$. The solid and dashed curves refer to cases where the calculation of the elastic strain energy measures $\gamma_\text{el}$ and $\Gamma_\text{el}$ is done based on the effective volume fraction $\kappa^\ast$ and nominal volume fraction $\kappa^0$, respectively, cf.~Eqs.~\eqref{Eq-adj} and \eqref{Eq-far}. Notice that, apart from the initial segment where $h$ is relatively small, the two curves exhibit a reasonably good agreement.}
\label{Fig-GammaSize}
\end{figure}

Fig.~\ref{Fig-GammaSize} provides a demonstration of the size effects in terms of the elastic micro-strain energy measures $\gamma_\text{el}$ and $\Gamma_\text{el}$. It can be observed that while $\gamma_\text{el}$ exhibits a roughly linearly increasing trend, the energy factor $\Gamma_\text{el}$ exhibits a nonlinearly decreasing trend, seemingly an opposing behaviour to that of the excess interfacial energy density $\gamma_\text{int}^\text{exs}$ shown in Fig.~\ref{Fig-excessSize}(b). The diminishing trend of $\Gamma_\text{el}$ is a clear indication of the shift in the balance of energy as $h$ varies and highlights the interplay between the interfacial energy and elastic strain energy contributions. Although $\Gamma_\text{el}$ appears to converge towards a limit value, similar to other curves in Fig.~\ref{Fig-excessSize}, this limit value remains unattainable within the range of $h$ explored in this study. 

It is worth highlighting that the range of the values of the energy factor $\Gamma_\text{el}$ observed in Fig.~\ref{Fig-GammaSize}, from 2 MJ/m$^3$ to nearly 4 MJ/m$^3$, is consistent with the findings of Seiner et al.\ \citep{seiner2011finite}. Specifically, their `optimal' microstructure exhibits a value of 2.9 MJ/m$^3$, and their `experimental' microstructure exhibits a value of 4 MJ/m$^3$ (we have determined these values based on the overall elastic strain energy and domain geometry reported therein). {Indeed, upon extrapolating the simulation data to encompass larger twin spacings it becomes evident that our energy factor $\Gamma_\text{el}$ has a limit value of only marginally lower than 2 MJ/m$^3$. This substantiates the relevance of the quantitative comparison made with the data of \citep{seiner2011finite} in which a twin spacing of 10 $\mu$m was used.}

\begin{figure}
\centering
\hspace*{-1.7cm}
\includegraphics[width=1.2\textwidth]{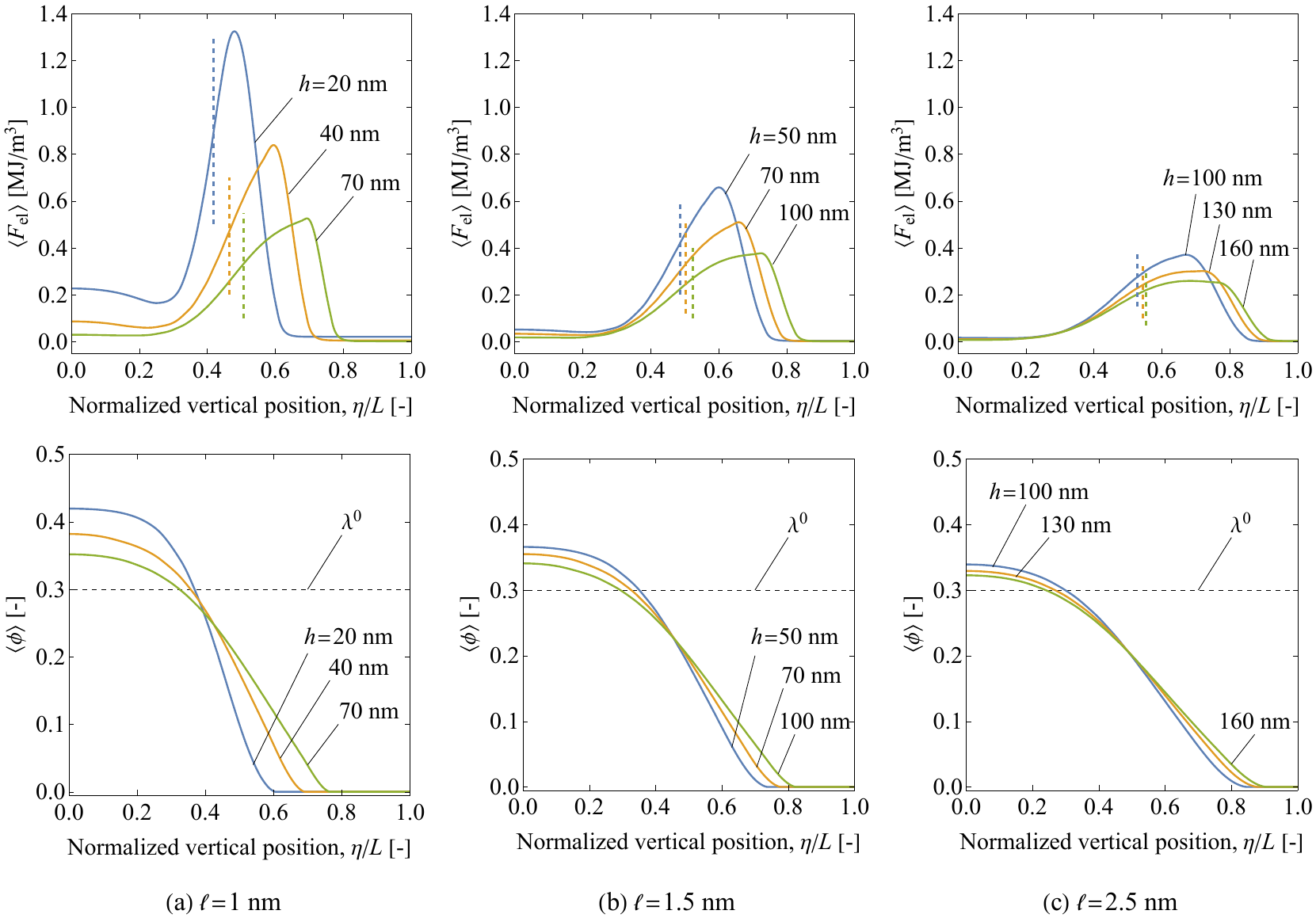}
\caption{The impact of twin spacing $h$ on the profile of the average elastic strain energy $\langle F_\text{el} \rangle$ (first row) and average order parameter $\langle \phi \rangle$ (second row) for three different interface thickness parameters $\ell$, namely (a) $\ell=1$ nm, (b) $\ell=1.5$ nm, and (c) $\ell=2.5$ nm. The vertical dashed lines in the first row indicate the position of the actual M--MM interface, which is specified in terms of the effective volume fraction $\kappa^\ast$.}
\label{Fig-ElasticPhi}
\end{figure}

In Figs.~\ref{Fig-excessSize} and \ref{Fig-GammaSize}, we have presented the collective responses showing the size effects. In order to gain a deeper understanding on the effect of twin spacing $h$, representative individual profiles of average elastic strain energy $\langle F_\text{el} \rangle$ and order parameter $\langle \phi \rangle$ are reported in Fig.~\ref{Fig-ElasticPhi}.

\begin{figure}
\centering
\includegraphics[width=1\textwidth]{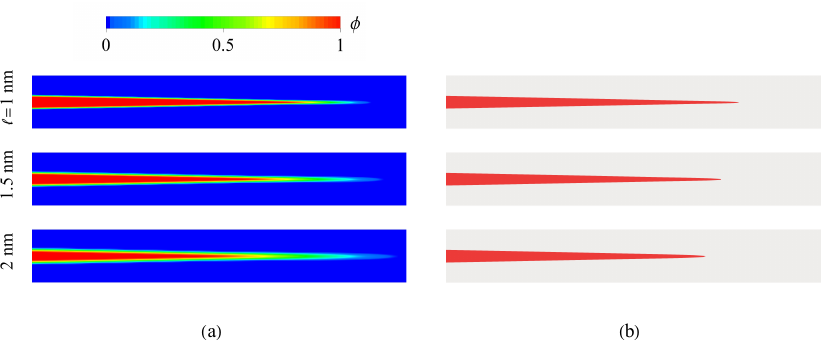}
\caption{Magnified views of the needle for varying interface thickness parameter $\ell$, with a fixed computational domain of $h \times L = 70 \times 1400$ nm. {In panel (b), the microstructures are represented by the trimmed order parameter $\phi \geq 0.5$ displayed by a single color, thus excluding the diffuse interfaces}.}
\label{Fig-NeedleTip}
\end{figure}

This section concludes with a discussion of the impact of the interface thickness parameter $\ell$ on the simulation results, as it is essential to ensure that the choice of $\ell$ does not compromise the validity of the analysis outcomes. While this can be partly discerned from Figs.~\ref{Fig-excessSize} and \ref{Fig-GammaSize}, further investigation deems necessary. Fig.~\ref{Fig-NeedleTip} presents magnified views of the needle-shaped martensite domains obtained for a fixed twin spacing $h=70$ nm but for various interface thickness parameters $\ell$. As expected, the apex of the needle becomes more diffuse as $\ell$ increases (Fig.~\ref{Fig-NeedleTip}(a)). However, no significant morphological changes are evident. This is also confirmed by the magnified views of the corresponding trimmed microstructure visualizations (Fig.~\ref{Fig-NeedleTip}(b)), in which the diffuse interfaces are excluded by representing variant B via trimmed order parameter $\phi \geq 0.5$ {and displaying it using a single color}.

\begin{figure}
\centering
\hspace*{-1.7cm}
\includegraphics[width=1.20\textwidth]{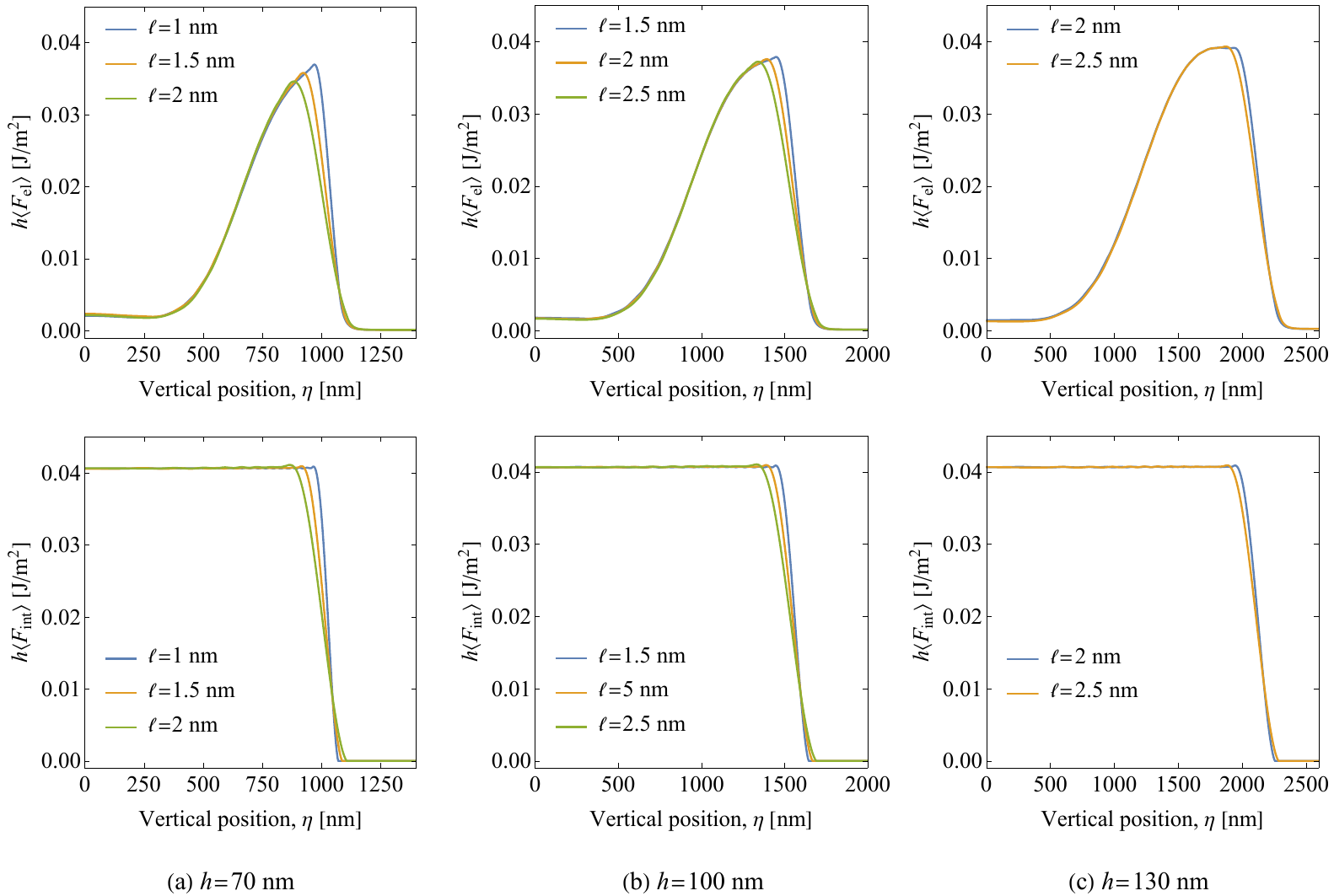}
\caption{The effect of interface thickness parameter $\ell$ on the profile of the elastic strain energy $h \langle F_\text{el} \rangle$ (first row) and on the profile of the interfacial energy $h \langle F_\text{int} \rangle$ (second row) for different twin spacings, namely (a) $h=70$ nm, (b) $h=100$ nm, and (c) $h=130$ nm.}
\label{Fig-ell}
\end{figure}

Furthermore, Fig.~\ref{Fig-ell} depicts the individual profiles of the elastic strain energy $h \langle F_\text{el} \rangle$ and interfacial energy $h \langle F_\text{int} \rangle$ for various $\ell$. Except for some visible effects in the vicinity of the energy peaks, which is expected due to the change in the diffuseness of the needle apex, the profiles are practically insensitive to the choice of $\ell$. It should be mentioned  that the same conclusion holds for the profile of the order parameter $\langle \phi \rangle$, which is not included in Fig.~\ref{Fig-ell}.

\subsection{Effect of twin volume fraction}\label{sec-vol}
The twin volume fraction $\lambda^0$ is regarded as a crucial input parameter in the analysis of the M--MM interface, as its selection has profound implications on the microstructure and the related quantitative characteristics. With that in mind, in this section, we seek to gain insight into the effect of twin volume fraction $\lambda^0$ on the simulation outcomes. To this end, simulations are conducted by varying $\lambda^0$ within the range of $0.1$ and $0.9$ with an increment of 0.1. A computational domain with dimensions of $h \times L=50 \times 1250$ nm$^2$ is selected, a nominal volume fraction of $\kappa^0=0.2$ is adopted and an interface thickness parameter of $\ell=0.5$ nm. The latter yields microstructures with less diffuse interfaces compared to those presented in preceding sections. To maintain the same microstructure resolution as before, i.e., to keep the ratio of $\pi \ell/d=5$, without a significant increase of the computational cost, a non-uniform finite-element mesh (non-uniform only in the longitudinal direction) is employed. More specifically, the mesh is finer in the vicinity of the needle apex (where the microstructure is more susceptible to morphological changes) and comprises nearly equiaxed elements of the size $d=0.3125$ nm, while it is coarser sufficiently far from the needle apex and comprises elongated elements. It should be remarked that our primary observation was that the morphology of the microstructure changes considerably within the $\lambda^0$ range of 0.5 to 0.7. Consequently, to augment the analysis, two additional simulations are performed for $\lambda^0=0.55$ and $\lambda^0=0.65$.

\begin{figure}
\centering
\hspace*{-0.5cm}
\includegraphics[width=1.1\textwidth]{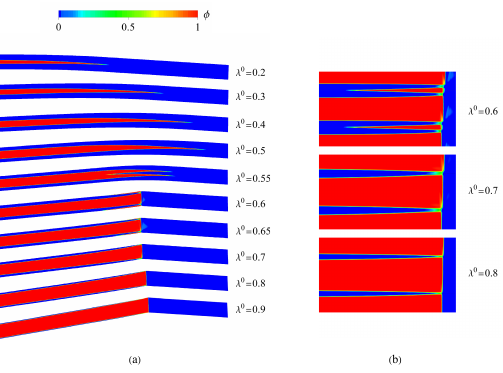}
\caption{The steady-state microstructure for various nominal twin volume fractions $\lambda^0$: (a) the microstructures in the deformed configuration for $\lambda^0$ values ranging from 0.2 to 0.9, and (b) magnified views of the periodically repeated microstructures (in the undeformed configuration) for $\lambda^0$ values of 0.6, 0.7 and 0.8. The microstructures for $\lambda^0=0.6$ and $\lambda^0=0.65$ exhibit some excessive diffuseness close to the needle apex, which can be circumvented by using a smaller interface thickness parameter $\ell$. Notice that, for space reasons, a portion of the microstructures in panel (a) is clipped from the left. As such, the microstructure for $\lambda^0=0.1$, which possesses a relatively shorter needle but otherwise is similar to that for $\lambda^0=0.2$, has not been included.}
\label{Fig-vol}
\end{figure}

The steady-state microstructures for different twin volume fractions $\lambda^0$ are compared in Fig.~\ref{Fig-vol}. Notably, two distinct families of microstructures emerge within the range of $\lambda^0$ investigated. For $\lambda^0 \leq 0.5$, the microstructure exhibits a single needle of variant B, which increases in height with increasing $\lambda^0$. On the other hand, for $\lambda^0 \geq 0.6$, variant A is engaged with the needle-shaped morphology and this is accompanied by the formation of an apparent interface between the domains of twinned martensite and single martensite. Unlike the microstructures for $\lambda^0 \leq 0.5$, the height of the needle is only minimally influenced by $\lambda^0$. An intriguing observation is that the two families of microstructures are mediated by a transitional microstructure at $\lambda^0=0.55$, which is clearly distinct from the microstructure of either family. Specifically, in this transitional microstructure, the needle-shaped domain of variant B exhibits a branching morphology, which arises as a means to reduce the elastic strain energy of the system. Similar observations of branching, as a spontaneous energy-minimizing mechanism, have been made in other phase-field modeling investigations \citep{finel2010phase,tuuma2016phase}. A closer examination of the microstructures for $\lambda^0=0.6$ and $\lambda^0=0.65$ reveals the occurrence of branching also in these cases, as can be seen clearly in the periodically repeated microstructure for $\lambda^0=0.6$ in Fig.~\ref{Fig-vol}(b). Another noteworthy observation from the microstructure visualizations in Fig.~\ref{Fig-vol}(b) is that the apparent interface between the twinned martensite and single martensite domains is not perfectly straight and takes on a stepped appearance. 

\begin{figure}
\centering
\hspace*{-1.5cm}
\includegraphics[width=1.15\textwidth]{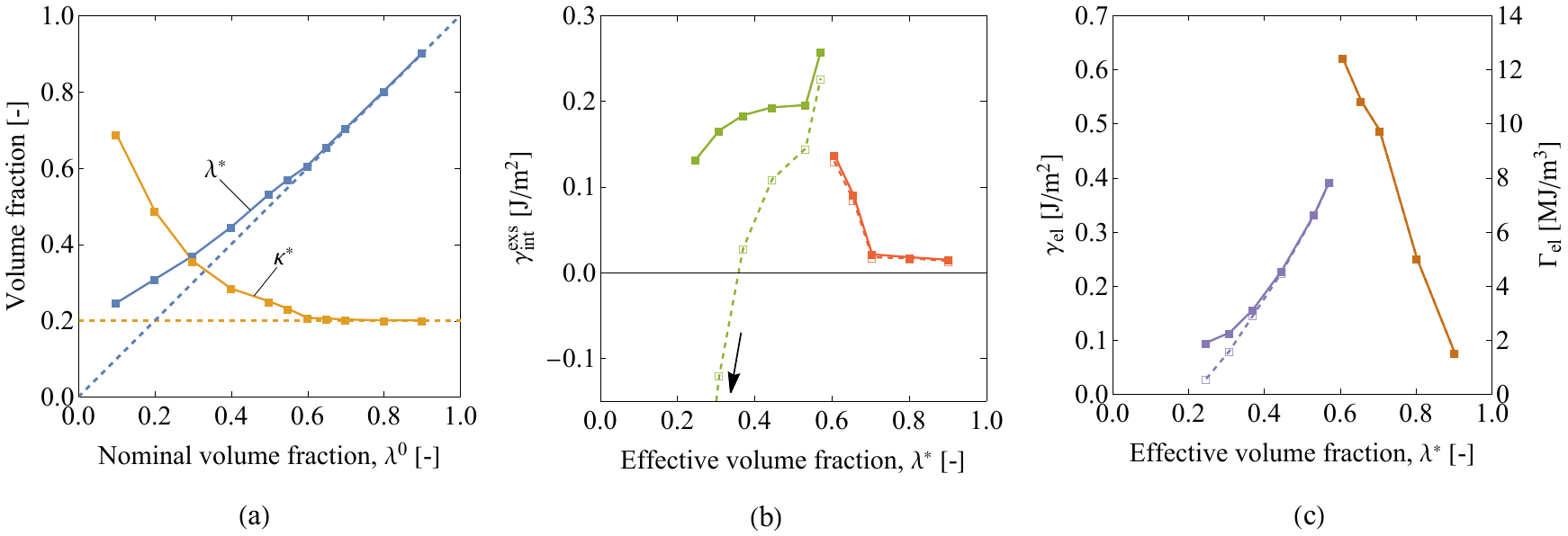}
\caption{The effect of the twin volume fraction $\lambda^0$ on (a) the effective volume fractions $\lambda^*$ and $\kappa^*$, (b) the excess interfacial energy density $\gamma_\text{int}^\text{exs}$, and (c) the elastic micro-strain energy measures $\gamma_\text{el}$ and $\Gamma_\text{el}$. In panel (a), the dashed curves indicate the nominal volume fractions $\lambda^0$ and $\kappa^0$. Notice that, in panels (b) and (c), the energy measures are plotted as a function of the effective volume fraction $\lambda^*$. Moreover, the solid and dashed curves refer to cases where the calculations are done based on the effective volume fraction $\kappa^*$ and nominal volume fraction $\kappa^0$, respectively.}
\label{Fig-GammaVol}
\end{figure}

Fig.~\ref{Fig-GammaVol} summarizes the effect of the twin volume fraction $\lambda^0$ on the macroscopic characteristics of the M--MM interface. Here, the primary observation pertains to the fact that the nominal and effective volume fractions tend towards each other as $\lambda^0$ increases, see Fig.~\ref{Fig-GammaVol}(a). In particular, for the second family of microstructures, i.e., for $\lambda^0 \geq 0.6$, the graphs of the effective and nominal volume fractions almost overlap. For the first family of microstructures, i.e., for $\lambda^0 \leq 0.5$, however, there exists a visible discrepancy (as already discussed in Section~\ref{sec-size} for $\lambda^0=0.3$) which diminishes as $\lambda^0$ increases. Consequently, the graphs of the excess interfacial energy density $\gamma_\text{int}^\text{exs}$ and elastic micro-strain energy $\gamma_\text{el}$ (and thus also $\Gamma_\text{el}$) (Fig.~\ref{Fig-GammaVol}(b,c)) exhibit distinctive behaviours for the two families of microstructures, and thereby, have been depicted by separate curves. Both $\gamma_\text{int}^\text{exs}$ and $\gamma_\text{el}$ energy measures have a monotonically increasing trend from both the right and left directions, i.e., as $\lambda^0$ increases for the first family and as $\lambda^0$ decreases for the second family, and peak at specific $\lambda^0$. More precisely, the peak for excess interfacial energy density $\gamma_\text{int}^\text{exs}$ occurs at $\lambda^0=0.55$, associated with the special branching morphology observed, and the peak for elastic micro-strain energy $\gamma_\text{el}$ occurs at $\lambda^0=0.6$.

It is worth mentioning that the graph of the elastic micro-strain energy $\gamma_\text{el}$ resembles qualitatively the bell-shaped curve proposed by Petryk et al.\ \citep{petryk2010interfacial}, which describes the {elastic micro-strain energy} of a generic transition layer linking laminated and homogeneous half-spaces. Their bell-shaped curve exhibits a symmetry with respect to the volume fraction of 0.5. Here, such a symmetry is not observed as a result of the substantial differences in the microstructures of the two families.

\section{Conclusion}
We have employed a conventional phase-field approach to model the microstructural features of the transition layer between a single martensite variant and a twinned martensite domain in a CuAlNi single crystal. The most salient feature of the microstructure is the presence of needle-like twins terminating at the interface, which has been successfully reproduced in our simulations, especially the bending and tapering of the needles that are in qualitative agreement with the experimental findings of Seiner et al.\ \citep{seiner2011finite}. Our primary objective has been to quantify the energy-based characteristics of the transition layer. In view of this, we have investigated the influence of the twin spacing (size effects) and the twin volume fraction on the interfacial and elastic strain energy measures. The obtained values, particularly for the elastic strain energy factor $\Gamma_\text{el}$ and the stresses, are in a quantitative agreement with those obtained in \citep{seiner2011finite} using a sharp-interface approach. A notable outcome of our analysis is the emergence of branching microstructure for certain twin volume fractions. Also, the microstructures and the energy measures predicted for type-I and type-II twins are found to be surprisingly similar. Our study exhibited a small impact of the phase-field length-scale parameter on the simulation results, confirming the validity of our analysis outcomes. 

\paragraph{Acknowledgement} This research was funded in part by the National Science Centre (NCN) in Poland through the Grant No.\ 2021/43/D/ST8/02555. For the purpose of Open Access, the authors have applied a CC-BY public copyright license to any Author Accepted Manuscript (AAM) version arising from this submission.

\appendix
\section{Effect of twin type}\label{sec-twintype}
\begin{figure}
\centering
\hspace*{-1.5cm}
\includegraphics[width=1.15\textwidth]{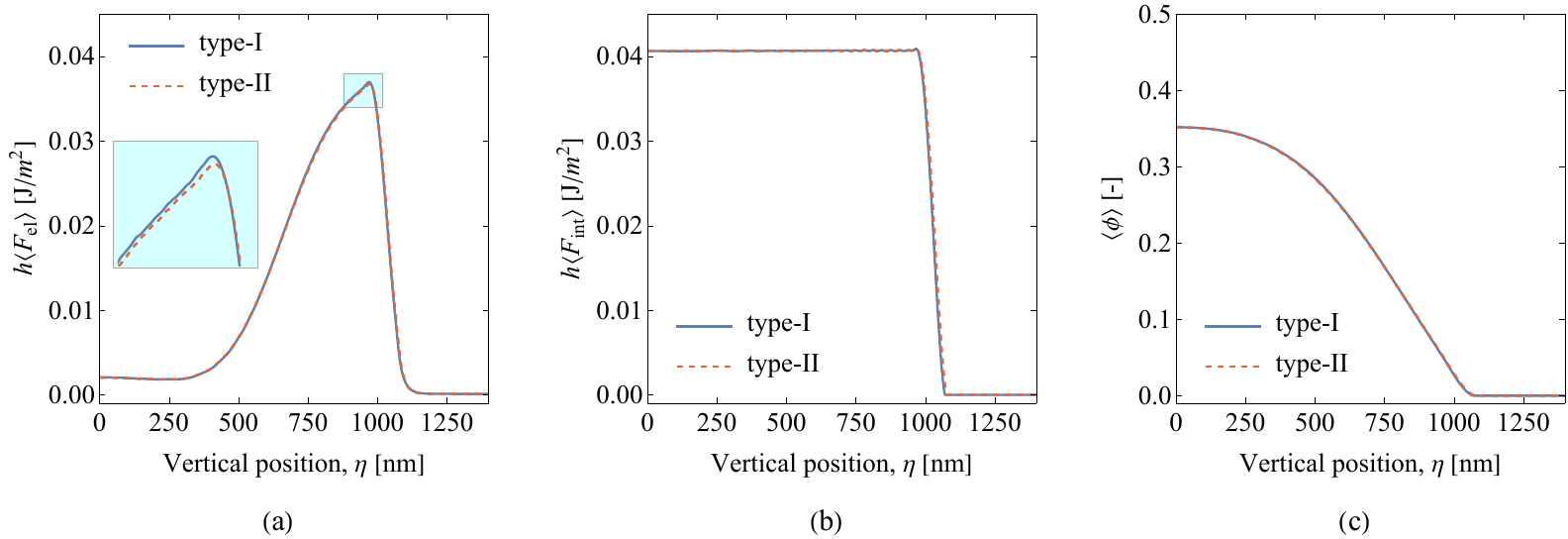}
\caption{The effect of the twin type on the simulation results for the representative study ($\lambda^0=0.3$, see Section~\ref{sec-rep}): (a) the profile of the elastic strain energy $h \langle F_\text{el} \rangle$, (b) the profile of the interfacial energy $h \langle F_\text{int} \rangle$, and (c) the profile of the average order parameter $\langle \phi \rangle$.}
\label{Fig-reptype}
\end{figure}

\begin{figure}
\centering
\hspace*{-1.8cm}
\includegraphics[width=1.2\textwidth]{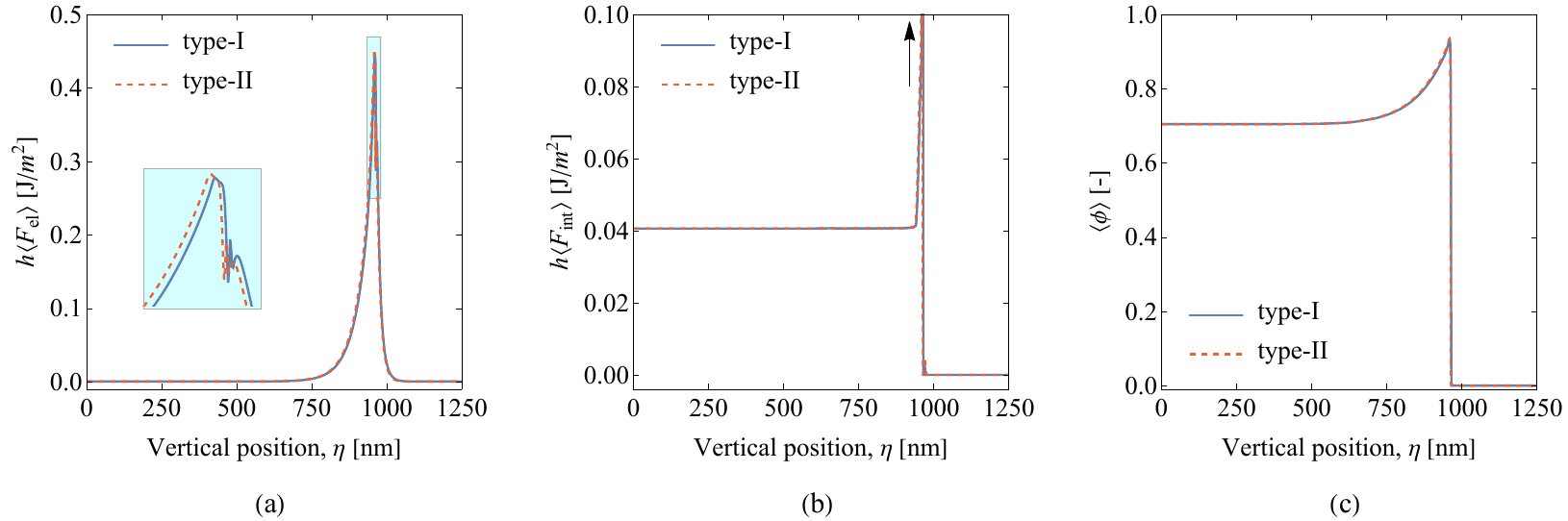}
\caption{The effect of the twin type on the simulation results for the case with the nominal volume fraction $\lambda^0=0.7$ (see Section~\ref{sec-vol}): (a) the profile of the elastic strain energy $h \langle F_\text{el} \rangle$, (b) the profile of the interfacial energy $h \langle F_\text{int} \rangle$, and (c) the profile of the average order parameter $\langle \phi \rangle$.}
\label{Fig-voltype}
\end{figure}

In this appendix, we present the simulation results obtained for type-II twin and compare them with those of type-I twin. Two cases are selected for this analysis, namely the case with the dimensions of $h \times L=70 \times 1400$ mm$^2$ and the nominal volume fractions of $\kappa^0=0.4$ and $\lambda^0=0.3$ (the representative case discussed in Section~\ref{sec-rep}) and the case with the dimensions of $h \times L=50 \times 1250$ and the nominal volume fractions of $\kappa^0=0.2$ and $\lambda^0=0.7$ (see the study of the effect of twin volume fraction in Section~\ref{sec-vol}). The longitudinal profiles of the integrated elastic strain energy $h \langle F_\text{el} \rangle$, the integrated interfacial interfacial energy $h \langle F_\text{int} \rangle$ and the average order parameter $\langle \phi \rangle$ for the two twin types are compared in Figs.~\ref{Fig-reptype} and \ref{Fig-voltype}. The results consistently reveal that the effect of twin type is negligible, as only some truly minor discrepancies can be detected, see the insets in Figs.~\ref{Fig-reptype}(a) and \ref{Fig-voltype}(a).

{It is to be remarked that for these additional simulations, the rank-one mixing \eqref{Eq-rankone} is redefined in terms of the type-II solution of the twinning equation, i.e., $\bfm{F}^\text{t}=\bfm{U}_\text{A}+ \phi \bfm{a}^\ast \otimes \bfm{l}^\ast$, where $\bfm{a}^\ast=(-0.0036,0.1691,-0.1921)$ and $\bfm{l}^\ast=(0.2282,0.6885,0.6885)$.}

\bibliographystyle{elsarticle-num}
\biboptions{sort&compress}

\section*{Data availability}
Data will be made available on request.

\bibliography{bibliography}

\end{document}